# Incorporating an Optical Clock into a Time Scale

Jian Yao[1], Thomas Parker[2], Neil Ashby[2], and Judah Levine[1]

[1]Time and Frequency Division and JILA, National Institute of Standards and Technology (NIST) and University of Colorado, Boulder, Colorado 80305, USA

[2]Time and Frequency Division, National Institute of Standards and Technology (NIST), Boulder, Colorado 80305, USA

Email: jian.yao@nist.gov; judah.levine@nist.gov

**Abstract**

This paper discusses how to build a time scale with an intermittently-operated optical clock. In particular, it gives suggestions on how long and how often to run an optical clock. It also explores the benefits of having an optical clock in a time scale, by comparing with the current UTC(NIST) performance and the time scale with a continuously-operated Cs fountain.

**Key Words**

Time scale, optical clock, UTC(NIST), Cs fountain, hydrogen maser, UTC.

## I.   Introduction

Nowadays, there are a few types of high-performance optical clocks, such as the $Al^+$ optical clock [1], and the Yb/Sr lattice clock [2-3]. They show a stability and accuracy of $1\times10^{-16}$ at an hour, or even much better. However, due to engineering obstacles, it is challenging to run an optical clock continuously for a few days, so up to now, people are hesitant to include an optical clock in a time scale. Instead of waiting until optical clocks become more reliable and can operate for longer periods of time, this paper discusses how to incorporate an intermittently-operated optical clock into a time scale, which is a temporary but practical solution for the time being. It will show that one can benefit from an optical clock, even though the clock only occasionally runs for periods of a few hours.

Section II discusses the performance of the current UTC(NIST) and the performance of a typical Cs-fountain based time scale. Note, in this paper, a Cs-fountain time scale means a time scale composed of a continuously (or near continuously)-operated Cs fountain and a free-running Hydrogen maser (or a free-running Hydrogen maser ensemble) that is frequency locked to the Cs-fountain through a separate frequency synthesizer. Section III explores the performance of a time scale with an intermittently-operated optical clock. Especially, we want to know how long and how often to run an optical clock, in order to make the time scale better than a Cs-fountain time scale. Section IV discusses some issues in an optical-clock time scale and Section V concludes this paper.

## II. Performance of UTC(NIST) and Performance of a Cs-Fountain Time Scale

UTC(NIST), as an official time in the United States, is generated by steering the free-running time scale AT1 (or TSC, which is a backup free-running time scale at NIST) to the Coordinated Universal Time (UTC) [4]. In other words, UTC(NIST) is a local realization of UTC at NIST. The blue curve in Figure 1 shows the frequency stability of AT1, against UTC, during Modified Julian Date (MJD) 56649 – 57514. We can see that AT1 reaches a fractional frequency stability in the mid $10^{-16}$ at around 20 days and then it starts to randomly walk. Note, the time-transfer noise contributes to the stability of the AT1 curve at around 5 days.

The magenta curve in Figure 1 shows the frequency stability of UTC(NIST), against UTC-UTC, during MJD 56649 - 57514. Clearly, the frequency stability is significantly improved relative to AT1 after ~100 days since UTC(NIST) is steered to UTC. However, the mid-term performance (10 days to 50 days) is sacrificed. This is because the BIPM (International Bureau of Weights and Measures) does not publish the UTC in real-time [5]. Instead, there is a latency of about 30 days. This makes a bump at ~30 days in the magenta curve. The time difference of UTC – UTC(NIST), during MJD 56649 to 57514, is shown in Figure 2. We can see that UTC(NIST) can deviate from UTC by more than 10 ns.

From the above discussion, the real-time (or short-latency) estimation of the frequency and time error in UTC(NIST) is very critical in the generation of UTC(NIST). To provide a more-frequent version of UTC, the BIPM publishes the rapid UTC (UTCr) every week [5]. Thus, the latency decreases from ~30 days to ~7 days. In principle, we have a faster estimation of the frequency error in UTC(NIST), by comparing it with UTCr. This can eventually improve the performance of UTC(NIST). However, UTCr is noticeably noisier than UTC and we are often chasing the noise. Steering to UTCr does not provide an obvious improvement of UTC(NIST) (see Figure 3).

This suggests that to improve the performance of UTC(NIST), we need to have a more stable and/or accurate frequency standard (e.g., an optical clock or a continuously-operated Cs fountain) at NIST, so that we can estimate the frequency and frequency drift of UTC(NIST) in real-time or near real-time.

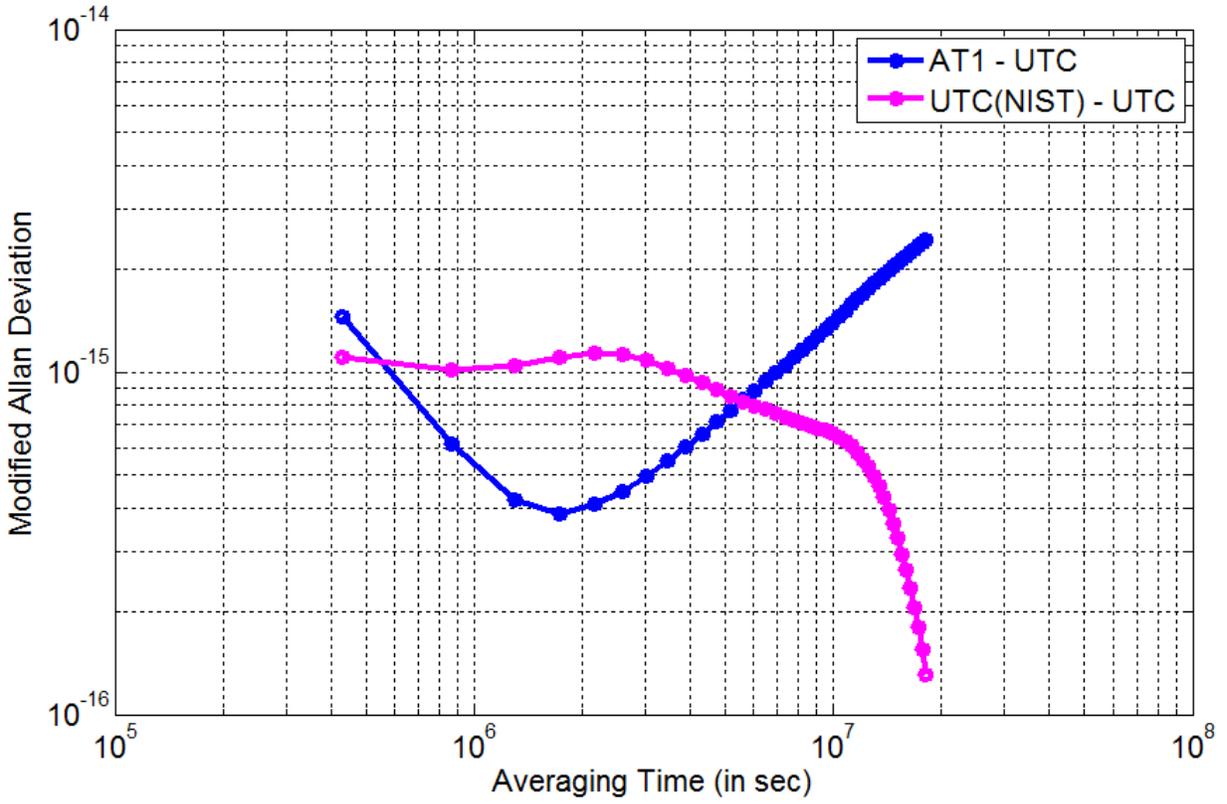

Figure 1. Performance of AT1 (blue curve) and UTC(NIST) (magenta curve), against UTC.

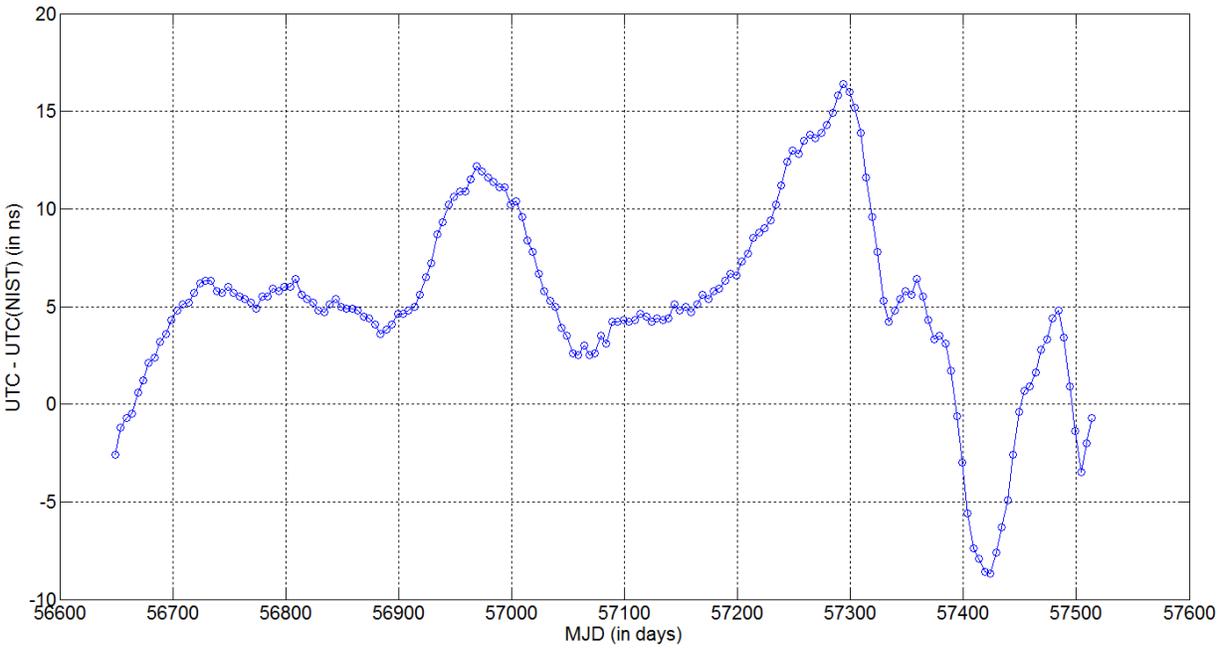

Figure 2. Time difference between UTC and UTC(NIST).

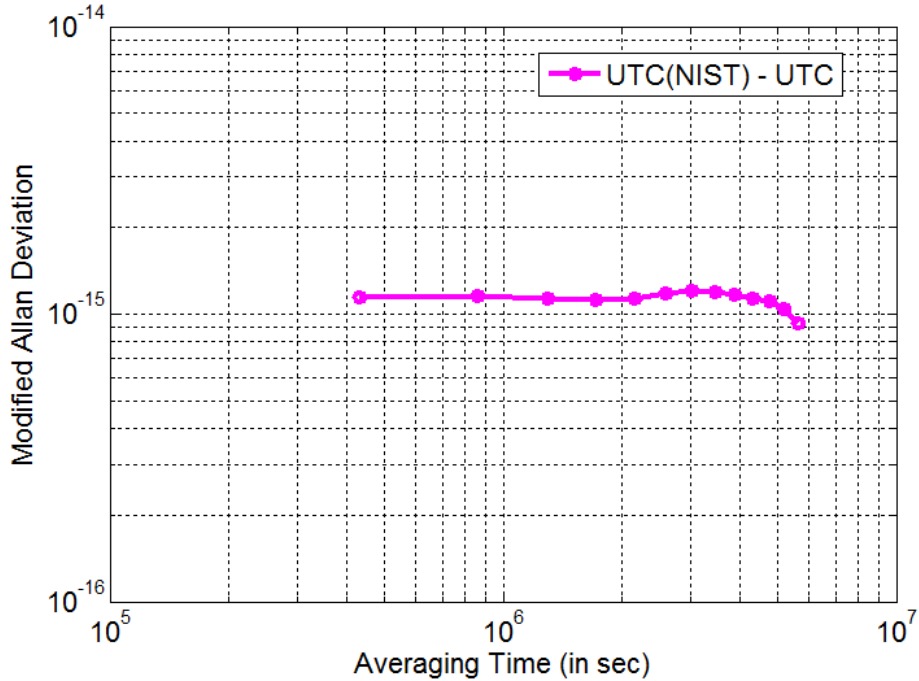

Figure 3. Performance of UTC(NIST) after steering closely to UTC_rapid. Note, we compute the frequency stability of UTC(NIST) for MJD 57504 to 57784, during which the UTC(NIST) is steered closely to UTC_rapid.

Next, we show the performance of a typical Cs-fountain time scale (i.e., a free-running hydrogen maser or hydrogen maser ensemble steered to a Cs fountain), by simulation. We assume the frequency stability of the continuously-operated Cs fountain to be approximately $4\times10^{-13}$ at 1 s and to be white FM in nature. We also simulate a free-running clock whose frequency stability is shown in Figure 4. This performance represents the performance of a typical hydrogen maser time scale or an excellent hydrogen maser with frequency drift essentially removed by modeling the drift of the maser, or time scale, with information from primary frequency standards. Then we steer the free-running clock to the Cs fountain, and thus form the Cs-fountain time scale.

The solid green curve in Figure 5 shows the frequency stability of the simulated Cs-fountain time scale. It has the short-term stability of the curve in Fig. 4 and the long-term stability of the Cs-fountain. Note, the magenta curve in Figure 5 is the same as the magenta curve in Figure 1, which shows the frequency stability of UTC(NIST) against UTC. The black and red curves show the frequency stability of UTC(OP)-UTC (the official time of France, generated at Observatoire de Paris) [6] and UTC(PTB)-UTC (the official time of Germany, generated at Physikalisch-Technische Bundesanstalt) [7] during MJD 56649 – 57514, respectively.

Note that OP and PTB both have continuously-operated Cs fountains. Therefore, UTC(OP) and UTC(PTB), as Cs-fountain time scales, should have a performance similar to the green curve. However, comparing the black/red curves with the green curve, we can see that the real Cs-fountain time scales are somewhat worse than the simulated Cs-fountain time scale for an averaging time of 5 – 150 days. This is because the noise in UTC itself contributes to the black/red curves at times less than about 100 days. We can also see that after ~ 150 days, there is a quick drop in the black/red curve, because UTC(OP) and UTC(PTB) are phase-locked to UTC.

For the magenta curve, it represents the frequency stability of UTC(NIST), since the noise of UTC is much smaller than that of UTC(NIST). Comparing the green solid curve with the magenta curve in Figure 5, we can see that the Cs-fountain time scale is much more stable than the UTC(NIST), in the long term. For example, the Cs-fountain time scale is more than twice as stable as UTC(NIST) at 5 to 20 days, and about five times as stable as UTC(NIST) at 100 days.

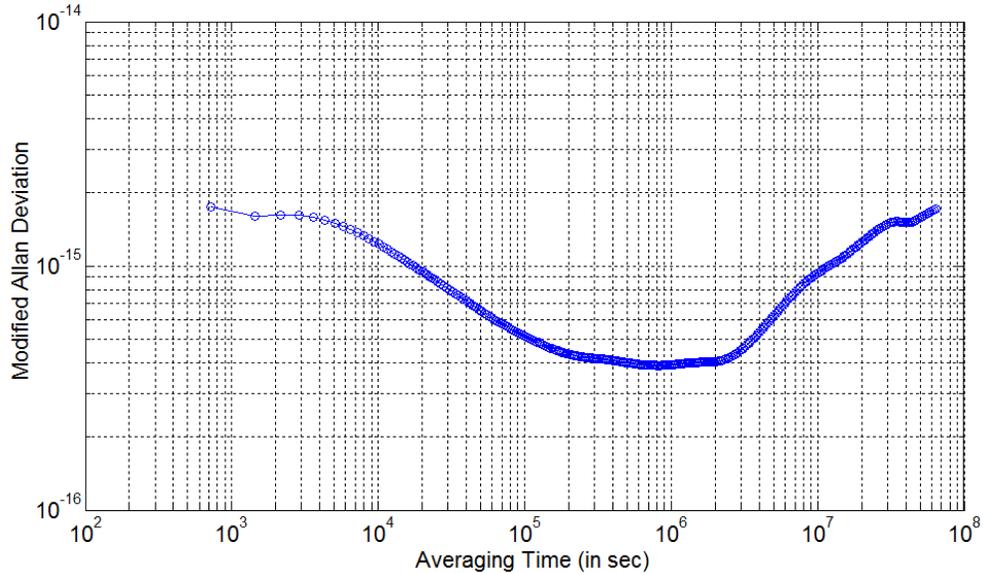

Figure 4. Performance of a typical free-running hydrogen maser time scale.

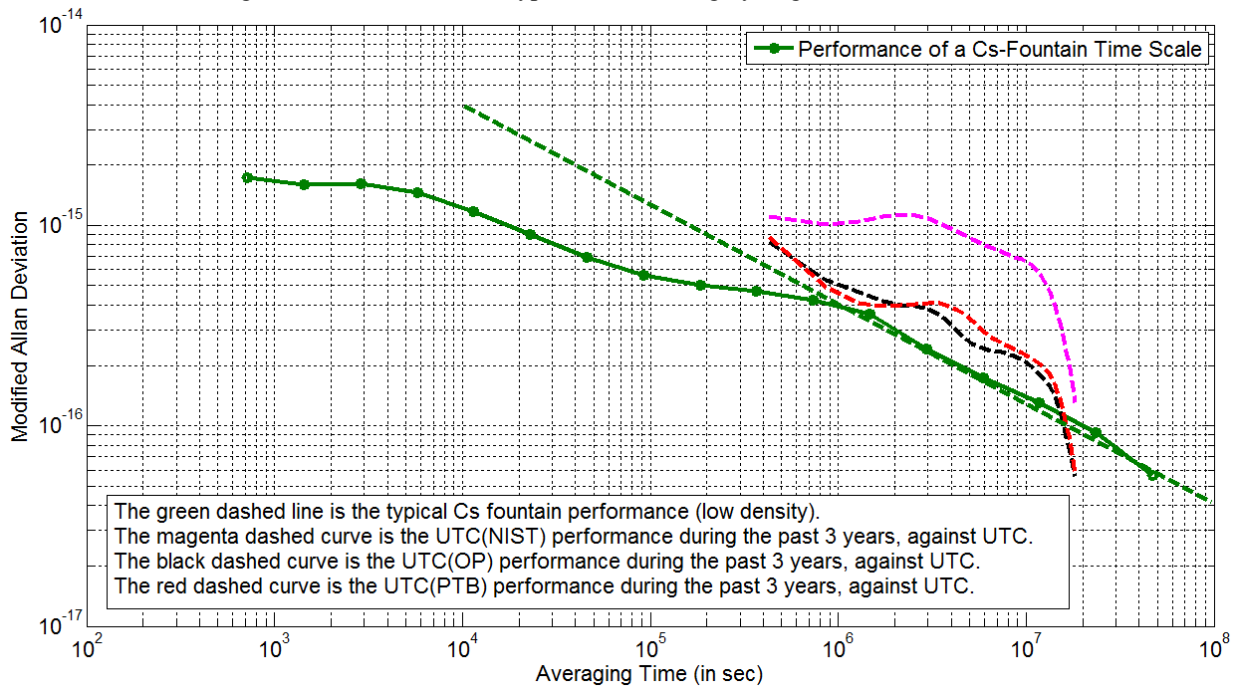

Figure 5. Performance of a Cs-fountain time scale (green solid curve). The magenta dashed curve, the black dashed curve, and the red dashed curve show the performance of UTC(NIST), UTC(OP), and UTC(PTB), against UTC, respectively. The green dashed line shows the performance of a typical continuously-operated Cs fountain.

## III. Performance of a Time Scale with an Intermittently-Operated Optical Clock

To build a time scale with an optical clock, we want it to be at least comparable to a Cs-fountain time scale. It is certainly good if an optical-clock time scale is significantly better than a Cs-fountain time scale. However, in practice, UTC cannot be better than $3.5\times10^{-16}$ because of its flicker floor [8]. Thus, it is not necessary to pursue a time scale significantly better than a Cs-fountain time scale for the time being. In the following, we explore how long and how often to run an optical clock so that the optical-clock time scale has a comparable performance to a Cs-fountain time scale.

First, we simulate a free-running maser-based time scale, whose frequency stability is already shown in Figure 4, with respect to an ideal time reference. Then we simulate an intermittently-operated optical clock with a frequency stability of $3.4\times10^{-16}$ at 1 s [3], and with white FM noise characteristics. Next, we steer the free-running time scale to the optical clock. The steering is done by estimating and then correcting the frequency and frequency drift of the free-running time scale, using a normal Kalman filter [9]. The benefit of using a Kalman filter is that this filter mathematically provides the optimal estimate. Also, it can automatically handle unevenly-spaced measurements. In our Kalman filter, we have optimized the parameters, such as the Q and R matrices, so that we get the best frequency stability. Last, the steered time scale is compared with the ideal time reference, and we compute the corresponding frequency stability.

For the intermittently-operated optical clock, it is expected that different operating schedules will give different levels of performance for the optical-clock time scale. In the following paragraphs, we study a variety of running schedules. This can provide guidance to people who run optical clocks.

Figure 6 shows the performance of the time scale with an optical clock running every half a day. We can see that by running an optical clock for 12 min every 12 hours, we can achieve the same performance as that of a continuously-operated Cs fountain time scale. Note, the operating time of the optical clock is only 1.667 %! We can also see that there seems to be little improvement if we run 24 min every 12 hours. This is due to the maser's flat noise characteristics at times on the order of 1000 s (see Figure 4).

Figure 7 shows the performance of the time scale with an optical clock running once a day. According to the red curve in Figure 7, we can achieve the same performance as that of a continuously-operated Cs-fountain time scale, by running an optical clock for 1 hour every day. We should emphasize that although only 12-min operation of an optical clock each day gives worse performance than the continuously-operated Cs fountain, it still generates a time scale which is significantly better than the current UTC(NIST). This can be seen by comparing the blue curve with the magenta dashed curve. Last, if we run an optical clock 12 hours a day, the frequency stability of the time scale will be improved to three times better than the Cs-fountain time scale, and about an order of magnitude better than the current UTC(NIST), for an averaging time of greater than 5 days. However, as we have mentioned earlier, this excellent performance is not necessary for the time being, since it is too good. Figure 8 shows the time series of the time scale with an optical clock running 1 hour a day. This provides an intuitive presentation of how much an optical-clock time scale varies from the UTC. In this case, we can see that the peak-to-peak time error of an optical-clock time scale is only about 2 ns, for 200 days. For longer than 200 days, we can phase lock the time scale to the UTC, so that the time error is always kept within 2 ns. Compared with the peak-to-peak time error of 25 ns in Figure 2, the improvement is significant.

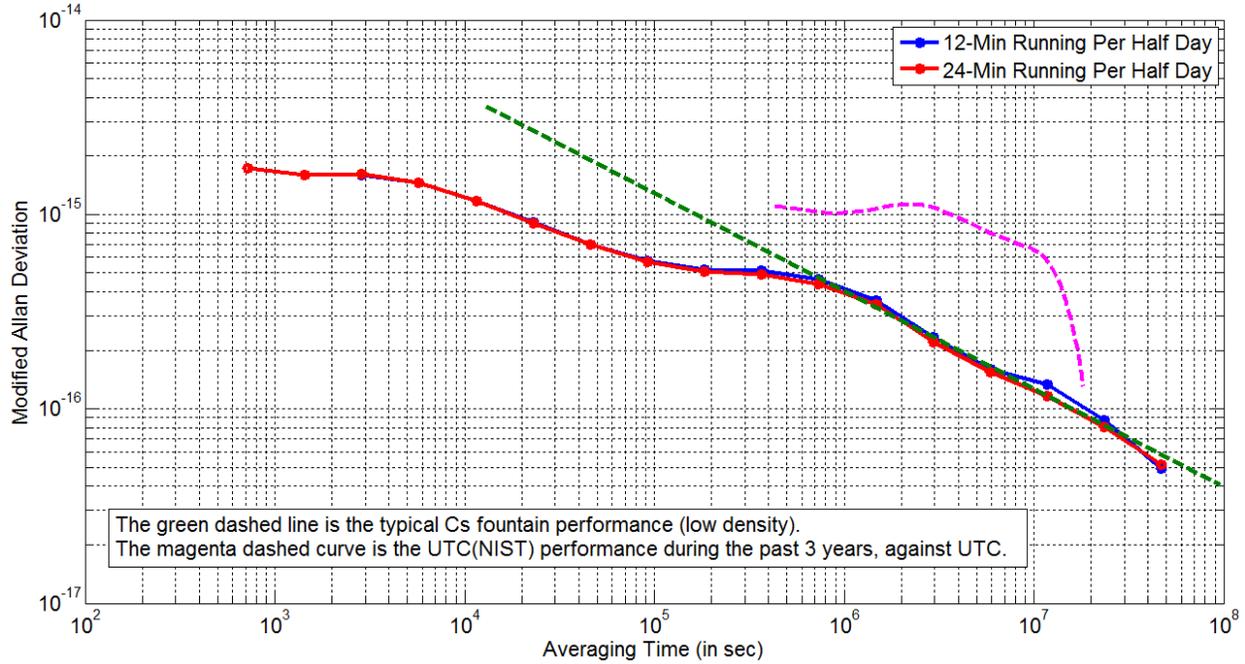

Figure 6. Running an optical clock every half a day.

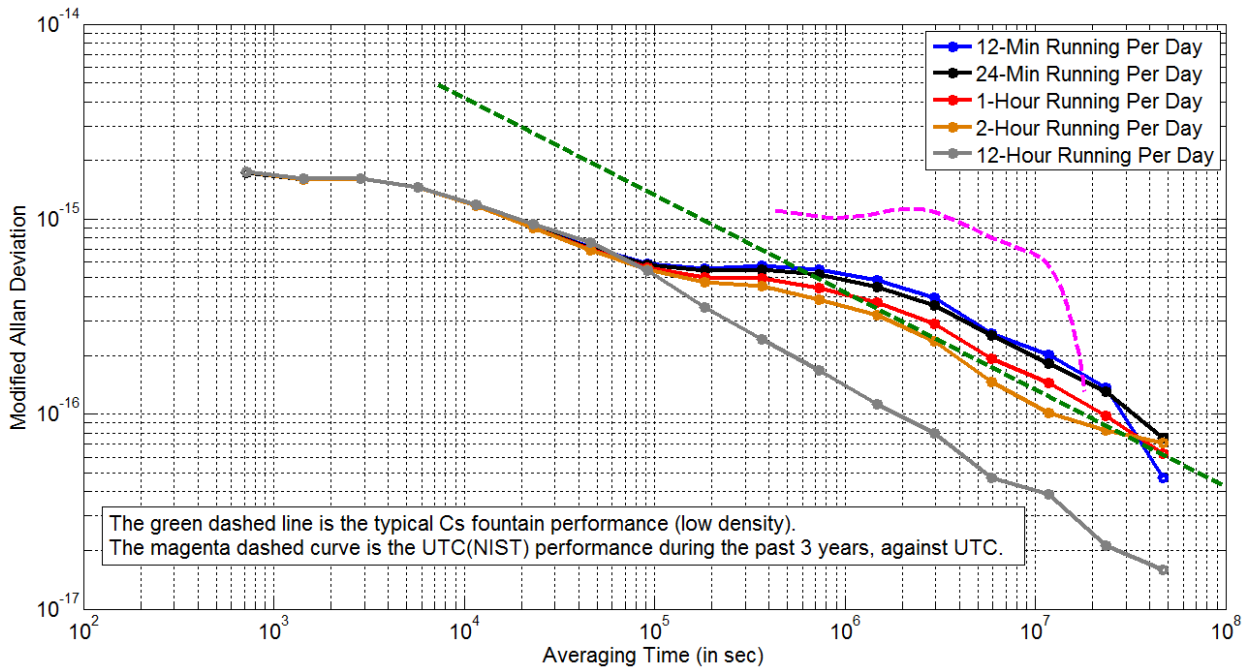

Figure 7. Running an optical clock once a day.

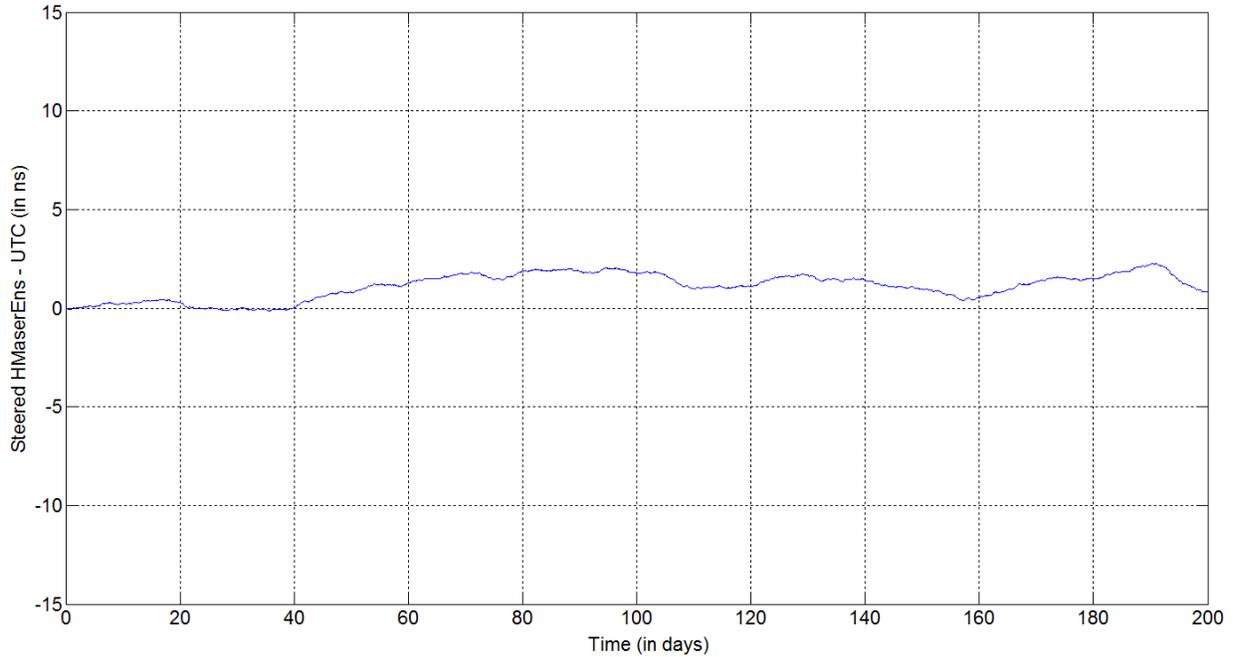

Figure 8. The time-domain performance of the time scale with an optical clock running 1 hour per day.

Figure 9 shows the performance of the time scale with an optical clock running three times every week. To achieve performance similar to that of a Cs-fountain time scale, we need to run an optical clock for at least four hours, three times a week.

Figure 10 shows the performance of the steered time scale with an optical clock running once a week. To achieve performance similar to that of a Cs-fountain time scale, we need to run an optical clock for at least 12 hours each week. From the black curve in Figure 10, even though a time scale with an optical clock running four hours a week is worse than a Cs-fountain time scale, the time scale is still much better than the current UTC(NIST). Even operating the optical clock 1 hour per week offers a modest improvement over the free running maser scale (see Fig. 4) at times larger than 20 days and could improve the performance of UTC(NIST).

From the above analysis, to achieve the same performance as a Cs-fountain time scale, we need to run an optical clock according to one of the following options: 12 min per half a day, 1 hour per day, 4 hours per 2.33 day, or 12 hours per week.

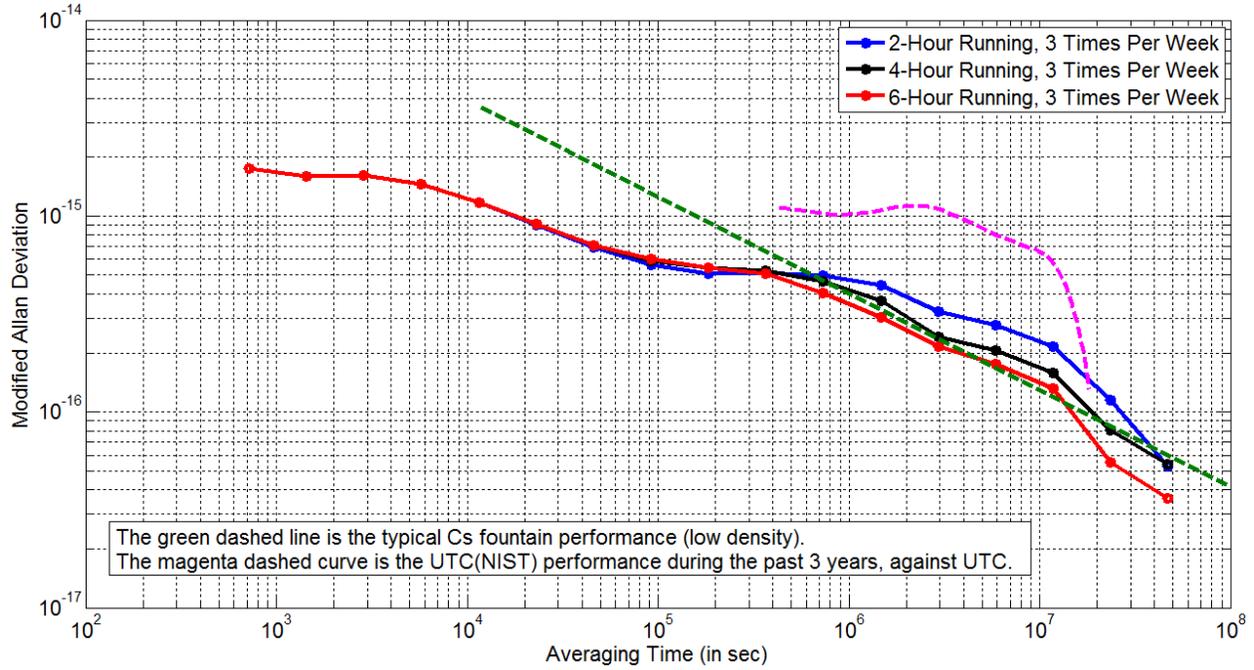

Figure 9. Running an optical clock 3 times a week.

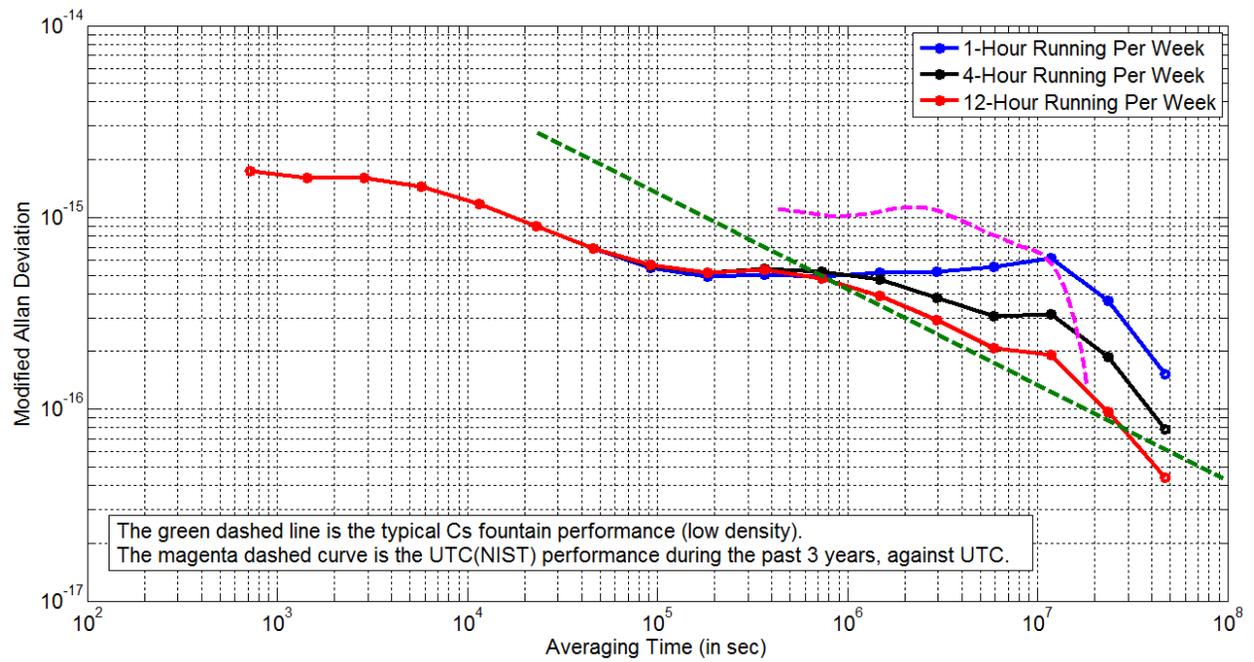

Figure 10. Running an optical clock once a week.

## IV. Discussions

Here, we discuss a few issues related to an optical-clock time scale.

First, as we know, a free-running time scale can be significantly pulled by a bad hydrogen maser, even though we monitor the hydrogen masers closely. The error becomes larger over time. Thus, a frequent evaluation of the free-running time scale is necessary to avoid a large timing error in UTC($k$) (i.e., a local realization of UTC at the timing laboratory "$k$"). Thus, to guarantee the quality of UTC($k$), we suggest running an optical clock three times a week, or even more frequently. Running an optical clock once a week may not be a good option.

Second, in Section III, we choose an optical clock with the white frequency noise of $3.4\times10^{-16}/\sqrt{\tau}$. If we use an optical clock with a worse performance (e.g., an Al$^+$ optical clock [1]), will our conclusions in Section III be the same? The answer is yes. This is because the free-running time scale is still much noisier than any optical clock. When we estimate the frequency of the time scale, the uncertainty in the estimation mainly comes from the noise in the time scale. Thus, any reasonable optical clock can fulfill the task of evaluating the time scale. Therefore, for a timing laboratory, we can have different optical clocks running at different time slots, assuming the accuracy is adequate. This can distribute the work to different groups of people and thus ease the work intensity.

Third, people are nowadays discussing steering a single hydrogen maser, instead of a time scale, to a continuously-running Cs/Rb fountain (the United States Naval Observatory, USNO, operates Rb fountains [10]). In this architecture, it seems that the only benefit of having a time scale is to improve the robustness of the timing system. However, in the architecture of an optical-clock time scale, using a time scale offers another benefit. That is, we can have a more accurate estimation of the frequency of the time scale when compared with an optical clock, because the short-term noise of a time scale is noticeably smaller than that of a single hydrogen maser due to averaging. Therefore, we can better correct the frequency error of the time scale, which eventually leads to an obvious improvement in the long-term (> 10 days) stability. In short, using a time scale, instead of a single hydrogen maser, provides a better long-term frequency stability of the optical-clock time scale.

## V. Conclusions

This paper studies how to build a time scale with an intermittently-operated optical clock. In particular, to achieve the same performance of a continuously-operated Cs-fountain time scale, we find that we need to run an optical clock 12 min per half a day, or 1 hour per day, or 4 hours per 2.33 day, or 12 hours per week. In the time series, the peak-to-peak timing error of an optical-clock time scale (e.g., run an optical clock 1 hour per day) is within 2 ns, as compared to the peak-to-peak timing error of 25 ns in the current UTC(NIST). In addition, even running an optical clock "12 min per day" would give an obvious improvement in UTC(NIST), at 5 to 150 days. These results are independent of the choice of an optical clock, as long as the optical clock's stability is well below that of the free-running time scale. The performance of the optical clock time scale depends on the inherent short-term stability of the microwave time scale. Thus, it is better to use a maser ensemble rather than a single maser.

ACKNOWLEDGMENTS

The clock noise in this paper is generated by the method proposed in [11]. We also thank Chris Oates, Jeff Sherman, Josh Savory, Victor Zhang, and Mike Lombardi for their helpful inputs.